\documentclass[reprint,amsmath,amssymb,aps,prl,twocolumn,superscriptaddress]{revtex4-1}
\usepackage{graphicx}
\usepackage{dcolumn}
\usepackage{bm}
\usepackage{color}

\begin{document}

\title{Management of the Correlations of Ultracold Bosons in Triple Wells} 
\author{Sunayana Dutta}
\affiliation{Department of Physics, Indian Institute of Technology  Guwahati, Guwahati-781039, Assam, India}
\author{Marios C. Tsatsos}
\affiliation{S$\tilde{a}$o Carlos Institute of Physics, University of S$\tilde{a}$o Paulo, P.O. Box 369, 13560-970 S$\tilde{a}$o Carlos, S$\tilde{a}$o Paulo, Brazil.}
\author{Saurabh Basu}
\affiliation{Department of Physics, Indian Institute of Technology  Guwahati, Guwahati-781039, Assam, India}
\author{Axel U. J. Lode}
\email{axel.lode@univie.ac.at}
\affiliation{Wolfgang Pauli Institute c/o Faculty of Mathematics, University of Vienna, Oskar-Morgenstern Platz 1, 1090 Vienna, Austria}
\affiliation{Vienna Center for Quantum Science and Technology, Atominstitut, TU Wien, Stadionallee 2, 1020 Vienna, Austria}

\date{\today}

\begin{abstract}
Ultracold interacting atoms are an excellent tool to study correlation functions of many-body systems that are generally eluding detection and manipulation. Herein, we investigate the ground state of bosons in a tilted triple-well potential and characterize the many-body state by the eigenvalues of its reduced one-body density matrix and Glauber correlation functions. We unveil how the interplay between the interaction strength and the tilt can be used to control the number of correlated wells as well as the fragmentation, i.e. the number of macroscopic eigenvalues of the reduced one-body density matrix.
\end{abstract}
\maketitle

\paragraph*{Introduction} The successful experimental realization of Bose-Einstein condensation in gases of ultracold rubidium atoms in periodic potentials, so-called optical lattices~\cite{Anderson1998,Greiner2001}, has provided a powerful platform to study numerous exotic quantum many-body phenomena~\cite{Gre,ek,dui}. The dimensionality and depth of the wells of the lattice can be experimentally tuned to control the configuration of particles. Remarkably, also the atom-atom interactions can be tuned via Feshbach resonances \cite{Ko,Blo,Chin}. 

Due to this impressive degree of experimental control, ultracold atoms in optical lattices can be used to mimic condensed matter systems and allow to simulate and probe their phase transitions~\cite{Bruder,Gre,Axel:17,esslinger_cqed,esslinger_ss}. Additionally, direct imaging of quantum many-body correlations is feasible: one-, two-, and even many-body correlations have already been detected~\cite{corr1,corr2,corr3,corr4}. 

Owing to its long decoherence time, the many-body state of ultracold atoms can provide a means to cache correlations and entanglement arising in quantum information processing~\cite{Bloch2008:Review}. For this purpose protocols to control and quantify correlations in the many-body state of ultracold atoms are necessary~\cite{Cramer2013}. Here, ultracold bosons in triple-well potentials provide a candidate system. Some of their many-body aspects have been previously studied~\cite{peters12,ultra,Stre}; however, a scheme to control the emergent correlations still needs to be devised.

We work out such a protocol for the management of correlations by including the tilt of the optical lattice as a control parameter. A \emph{tilted lattice} can routinely be achieved in the laboratory by superimposing a magnetic bias field to the optical potential. 
The inclusion of the tilt widens the spectrum of controllable parameters and enriches the emergent physics. For instance, Ising density wave order and the appearance of superfluidity in transverse directions of a system of ultracold charged bosons confined in a lattice with a tilt were described in Ref.~\cite{fifth}. Ref.~\cite{Hiller} demonstrates that some eigenstates in the spectrum of neutral bosons confined in a tilted one-dimensional lattice exhibit localization and are robust against external perturbations. Furthermore, Ref.~\cite{Fraze} shows that the tilt is a source of quantum decoherence for macroscopic quantum superpositions in ultracold atoms in a tilted well.

In this Letter, we study the many-body correlations in the ground state of interacting ultracold bosonic atoms in a tilted triple-well potential by solving the corresponding Schr\"{o}dinger equation using the multiconfigurational time-dependent Hartree for bosons (MCTDHB) approach.

We use the reduced density matrix (RDM) of the many-body state to quantify correlations.
The system is said to be coherent and condensed if only one eigenvalue of the RDM is macroscopic~\cite{nine} and is said to be correlated and fragmented if multiple eigenvalues of the RDM are macroscopic~\cite{Spekkens,Noizieres1982}. To get a spatially resolved understanding of the emergent correlations, we compute the Glauber first-order correlation function from the RDM~\cite{Glauber}.
We study the emergence of correlations and fragmentation in the many-body system as a function of the interaction strength and the tilt of the triple well. Our results unravel an intriguing interplay between the tilt of the lattice potential and the strength of the interparticle interactions. We demonstrate how this interplay can be exploited to manage the correlations and fragmentation of many-boson systems in tilted optical lattices to a large extent.

\paragraph*{Method} The properties of ultracold bosonic many-body systems are described by the time-dependent many-body Schr\"odinger equation for interacting and indistinguishable bosonic particles. Commonly, the many-body problem is solved by the mean-field Gross-Pitaevskii approximation~\cite{Peth,Pita} or the Bose-Hubbard model~\cite{Bruder,Gre}. In the Gross-Pitaevskii picture the RDM has only a single eigenvalue and hence correlations and fragmentation -- pivotal in the superfluid to Mott-insulator phase transition~\cite{Bruder,Gre} -- cannot be captured. In the Bose-Hubbard model a fixed basis set of Wannier states is utilized. Albeit being an apt choice for regular lattices, a Wannier basis may not be optimal for tilted lattices because the tilt renders the shape of the site-local single-particle states different from Wannier functions. 
The MCTDHB theory optimizes variationally both the basis set \emph{and} the expansion coefficients in that basis set (see~\cite{sec,third} and references therein); its solutions thus assume no predetermined symmetry or shape of the described many-body state. Therefore, we use MCTDHB to obtain an optimized problem-adapted basis to investigate tilted lattices. MCTDHB is in principle exact~\cite{Axel:Thesis,Axel:HIM}, can describe both coherent and fragmented condensates, and includes the GP theory as an extreme case when only one single-particle state is considered; see Refs.~\cite{third,Axel:Spin,Axel:MCTDHF} and Supplemental Material~\cite{SupplMat} for details on MCTDHB. 

The $N$-boson state $\vert \Psi \rangle$ is governed by the time-dependent Schr\"odinger equation 
\begin{equation}
 i \partial_t \vert \Psi \rangle = \hat{H} \vert \Psi \rangle, \label{TDSE}
\end{equation}
with the Hamiltonian
\begin{eqnarray}
\hat{H}(x_1,x_2,...,x_N) &=& \sum_{j=1}^{N} \hat{h}(x_j)+\sum_{k>j=1}^{N} \hat{W}(|x_j-x_k|).~~\label{Eq:H}
\end{eqnarray}
We compute the ground state of the Hamiltonian in Eq.~\eqref{Eq:H} by propagating Eq.~\eqref{TDSE} in imaginary time to damp out any excitation in the one-dimensional many-body system. 
Here, $x_{j}$ represents the position of the $j^{th}$ boson, $\hat{h}$ is the single-particle Hamiltonian $\hat{h}(x)=\hat{T}(x)+\hat{V}_{\text{trap}}(x)$; $\hat{T}(x)$ and $\hat{V}_{\text{trap}}(x)$ are the usual kinetic and external potential energy, respectively. Interactions of ultracold dilute bosonic gases are typically modeled using a Dirac-delta distribution: $\hat{W}(x_{j}-x_{k})=\lambda_{0} \delta(x_j - x_k)$. Here, $\lambda_{0}$ is referred to as the strength of interactions. We scale $\lambda_0$ with the particle number as $\lambda=\lambda_0(N-1)$. In Eqs.~\eqref{TDSE},\eqref{Eq:H} and the remainder of this work dimensionless units are employed~\cite{units}.

\begin{figure}[htbp]
\centering
\includegraphics[angle=-90,width=0.5\textwidth]{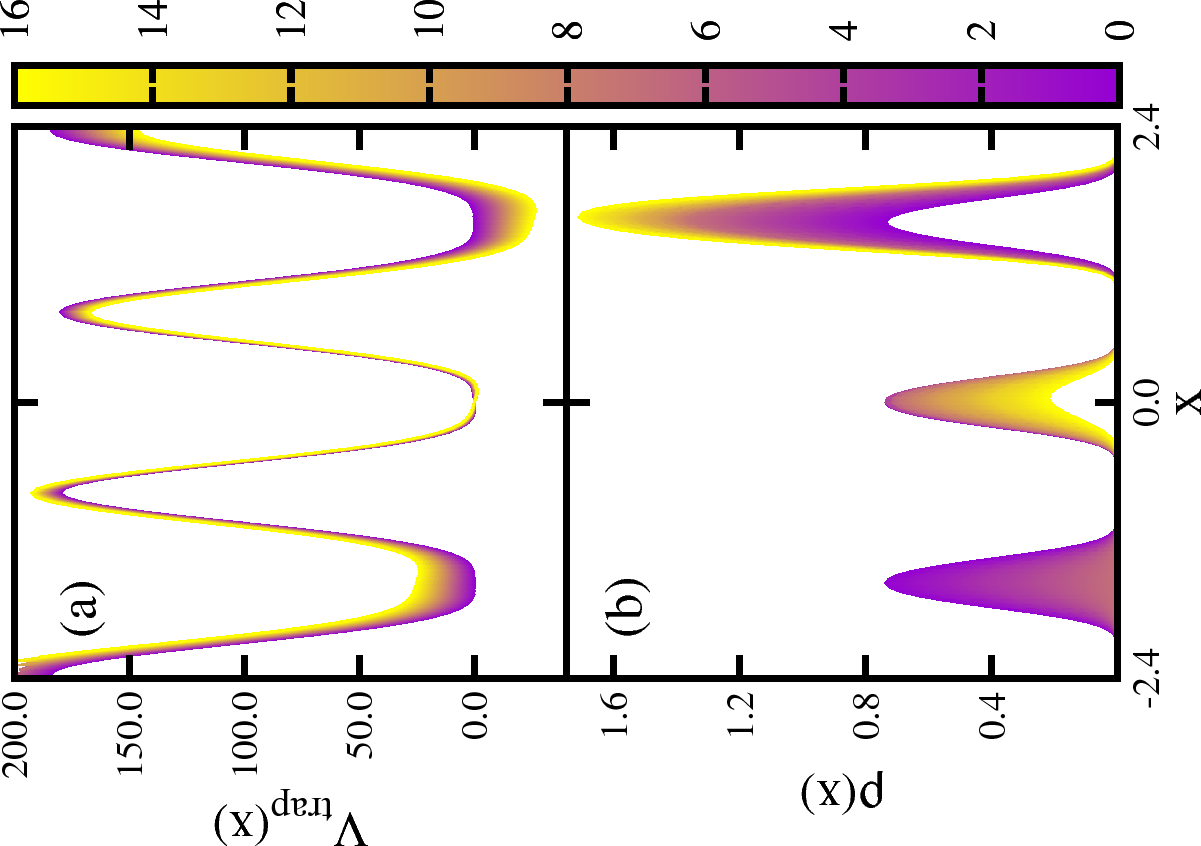}
\caption{Potential and density as a function of the tilt.
(a) shape of the triple well $V_{\text{trap}}(x)$ for $V_0=180$, $\lambda=6$ and various tilts $\alpha$ and (b) corresponding density $\rho(x)$ for the same $\alpha$, ranging from $\alpha=0$ to $\alpha=16$ (see color code/gray-scale). 
\label{Fig:1}}
\end{figure}

We consider $N=90$ interacting bosons in a trap of the form
\begin{equation}
V_{\text{trap}}(x) = -\alpha x+ V_{0}\sin^{4} (kx)+f_{w}(x). \label{Eq:Pot}
\end{equation}
Here $\alpha$ is the tilt and $V_{0}$ the barrier height. We fix $k=2$ for the lattice spacing. The term $f_{w}(x)$ introduces quasi-hard-wall boundary conditions~\cite{num_details}. The tilt ${\alpha x}$ renders the trapping potential similar to that of charged particles in a constant electric field and can be realized by applying a magnetic field gradient to ultracold neutral bosons in a lattice. The potential is plotted in Fig.~\ref{Fig:1}(a) for $\alpha\in[0,16]$.

The one-body reduced density matrix of the $N$-boson state $\vert \Psi(t)\rangle$ is defined as:
\begin{equation}
{\rho}^{(1)}(x,x') = \langle\Psi \vert \hat{\Psi}^\dagger(x') \hat{\Psi}(x) \vert \Psi \rangle =\sum_{i} n_{i}\phi_{i}^{*}(x')\phi_{i}(x), \label{Eq:RDM}
\end{equation}
in its eigenbasis $\{\phi_{i}(x)\}$ \cite{three,Ksak}.
Here $n_{i}$ is the $i^{th}$ eigenvalue and $\phi_{i}(x)$ the corresponding eigenfunction, also known as natural occupation and natural orbital, respectively. The diagonal of $\rho^{(1)}$ corresponds to the single-particle probability distribution $\rho(x)$. A BEC is condensed if its RDM has only a single macroscopic eigenvalue~\cite{nine} and $k$-fold fragmented, if its RDM has $k$ macroscopic eigenvalues. The first-order coherence of a condensed state is maintained everywhere in space. Therefore, the value of the first occupation $\frac{n_1}{N}\approx1$ ($\frac{n_1}{N}<1$) is also indicative of the (loss of) coherence of the state (see Eq.~\eqref{Eq:g1} below).

\paragraph*{Results} For our numerical calculations, we use $M=3$ one-dimensional single-particle basis functions and consider $N=90$ particles. We also tested $M>3$ for convergence; see Ref.~\cite{SupplMat}. For the present computations we use the MCTDH-X implementation of the MCTDHB theory~\cite{Axel:Spin,Axel:MCTDHF,package}.

We start our investigation by plotting the one-body density $\rho(x)$ in Fig.~\ref{Fig:1}(b) as a function of the tilt $\alpha$. The effect of the tilt on the density $\rho(x)$ is intuitive: as $\alpha$ increases the density of the atoms is gradually forced downhill and $\rho(x)$ is localized mostly at the rightmost well where the potential energy is minimal for $\alpha>0$.

We chose the values of the interaction strength ($\lambda=6$) and barrier height ($V_0=180$) such that the ground state is threefold fragmented in the absence of tilt ($\alpha=0$). To assess the impact of the barrier height and the interaction strength on the properties of the many-body state, we additionally consider a larger interaction strength, namely $\lambda=20$, and a moderate barrier height, namely $V_0=80$.

\begin{figure}[!]
\includegraphics [width=0.5\textwidth,angle=270]{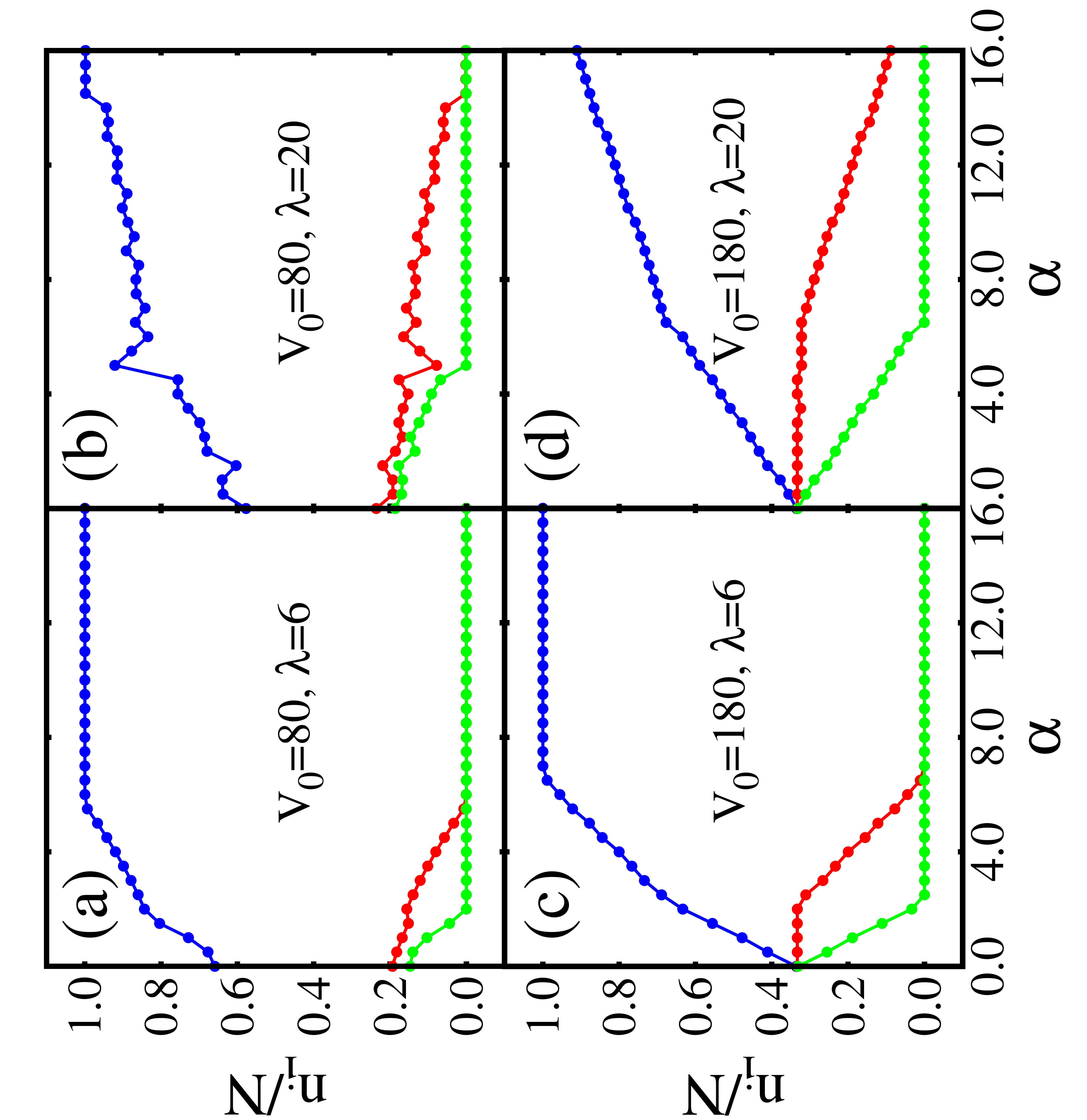}
\caption{Fragmentation as a function of the tilt of the lattice.
The natural occupations $\frac{n_{i}}{N}$ are shown as a function of the tilt $\alpha$ for barrier heights $V_0=80$ [$V_0=180$] in (a),(b) [(c),(d)]. Panels (a),(c) [(b),(d)] correspond to interaction strength $\lambda=6$ [$\lambda=20$]. In all panels, the blue line with circles represents $\frac{n_1}{N}$, the red line with circles represents $\frac{n_{2}}{N}$, and the green line with circles represents $\frac{n_{3}}{N}$. For all depicted parameters, fragmentation gradually diminishes with increasing tilt $\alpha$; for large tilts, the state hence becomes coherent and the occupation numbers obtained are $\frac{n_1}{N} \approx 1; \frac{n_2}{N}\approx \frac{n_3}{N} \approx 0$. All quantities shown are dimensionless, see text for further discussion.}
\label{Fig:2}
\end{figure}

To quantify the fragmentation, coherence, and correlation properties of the many-boson system we discuss the behavior of the natural occupations, $\frac{n_{i}}{N}$, as a function of the tilt $\alpha$ [cf. Eq.~\eqref{Eq:Pot}], see Fig.~\ref{Fig:2}.

For moderate barrier height, $V_{0}=80$, and no tilt, $\alpha=0$, the bosons are not completely fragmented, i.e., $\frac{n_1}{N}>60\%$ and $\frac{n_{2,3}}{N}<20\%$, for both small and large interaction strengths ($\lambda=6$ and $\lambda=20$). This is in contrast to the entirely threefold fragmented state found for $V_0=180$ with $\frac{n_1}{N}\approx\frac{n_2}{N}\approx\frac{n_3}{N}\approx33.33\%$ [cf. Fig.~\ref{Fig:2}, panels (a),(b) and Fig.~\ref{Fig:2}, panels (c),(d)]. We conclude that, at zero tilt, fragmentation can be tuned by the barrier height alone. As $\alpha$ grows larger so does the first natural occupation, $\frac{n_1}{N} \rightarrow 1$, while the other two natural occupations decrease, i.e. $\frac{n_{2,3}}{N} \rightarrow 0$, see Figs.~\ref{Fig:2}(a) and (b). 

At large barriers, $V_{0}=180$, and moderate interactions $\lambda=6$, the state exhibits threefold fragmentation at $\alpha=0$. As $\alpha$ increases past a threshold value of $\alpha \approx7.5$, the state becomes coherent with $\frac{n_1}{N}\approx 1,\frac{n_2}{N}\approx \frac{n_3}{N}\approx0$, see Fig.~\ref{Fig:2}(c). 
Interestingly, the second natural occupation $\frac{n_{2}}{N}$ remains constant up to tilts as large as $\alpha\approx2$, while $\frac{n_3}{N}$ starts to drop from $\frac{1}{3}$ to $0$ already at $\alpha \approx 0$. For $\alpha>2$, $n_2$ falls off gradually and vanishes at $\alpha\approx7.5$ [see Fig.~\ref{Fig:2}(c)]. Beyond this tilt the density is almost exclusively localized in the rightmost well. For larger barriers, $V_{0}=180$, and moderate interactions, $\lambda=6$, an increasing tilt $\alpha$ thus triggers a transition from a fully threefold fragmented to a fully condensed state, i.e., the tilt can be used to control fragmentation.

For larger interactions, $\lambda=20$, and a large barrier height, $V_0=180$, the transition between a fragmented and a depleted state is still found, however, at larger tilts $\alpha$ [compare Figs.~\ref{Fig:2}(c) and (d)]. 

We have verified that the above findings for the natural occupations and the fragmentation of the state also hold for the case of long-range interactions of the form $\hat W(r_j-r_k)=\lambda_0/(|x_j-x_k|^3+\Delta^3)$. The natural occupations follow the same pattern as their contact-interaction counterparts, but the restoration of coherence seems to happen at even larger values $\alpha$ as compared to the case of contact interactions. This demonstrates the sharper effect of long-range interactions on the fragmentation, see Supplemental Material~\cite{SupplMat}. We thus conclude that the tilt of the triple well can be used to tune the many-body state from fragmented to condensed.

To obtain a spatially resolved picture of the correlations between the atoms in the many-body state that are triggered by a specific trap geometry, we study the behavior of the first-order correlation function,
\begin{equation}
\vert g^{(1)}(x,x';t) \vert^2 = \left| \frac{\rho^{(1)}(x,x';t)}
{\sqrt{\rho^{(1)}(x,x;t)\rho^{(1)}(x',x';t)}} \right|^2. \label{Eq:g1}
\end{equation}

\begin{figure}[!t]
\hspace*{-0.9cm}
\includegraphics[width=0.48\textwidth,angle=-90]{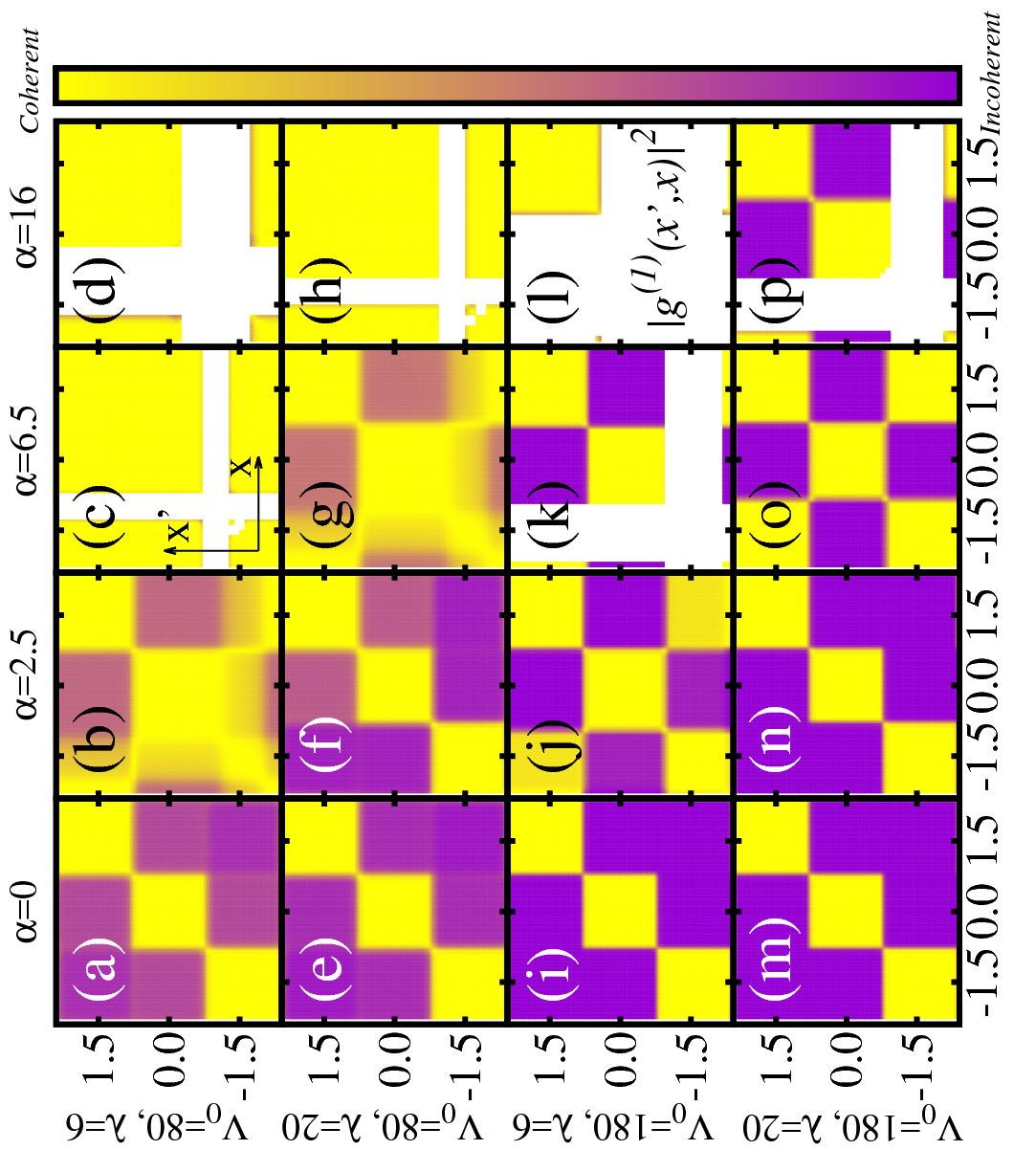}
\caption{Spatially tracing correlations between the bosons in the triple well as a function of the tilt and barrier height.
The first-order normalized correlation function $|g^{(1)}(x',x)|^{2}$ is visualized as a function of $\alpha$ wherever the density is larger than a threshold value, i.e., where $\rho^{(1)}(x,x)>0.01$ and $\rho^{(1)}(x',x')>0.01$. See labels for the respective values of the barrier height $V_0$ and the tilt $\alpha$ and text for discussion.
}
\label{Fig:3}
\end{figure}

The value $\vert g^{(1)}(x,x';t)\vert^2$ marks the first-order coherence between the points $x$ and $x'$ ($\vert g^{(1)}(x,x')\vert^2\approx1$) or its absence ($\vert g^{(1)}(x,x')\vert^2\approx0$) in the state $\vert \Psi \rangle$~\cite{Glauber}. In the following, we use the term \textit{inter-well coherence} if $x$ is in the vicinity of a different minimum of $V_{\text{trap}}$ than $x'$ and $\vert g^{(1)}(x,x')\vert^2\approx1$ holds. Moreover, we use the term \textit{intra-well coherence} if $\vert g^{(1)}(x,x')\vert^2\approx1$ holds for coordinates $x$ and $x'$ that are both in the vicinity of the same minimum.
We plot $\vert g^{(1)}\vert^2$ for various tilts ($\alpha=0,2.5,6.5,16$), barrier heights ($V_0=80,180$) and interaction strengths ($\lambda=6,20$) in Fig.~\ref{Fig:3}.

We first discuss the correlation function for a moderate barrier height, $V_{0}=80$, in Fig.~\ref{Fig:3}(a)--(h). 
At a small interaction strength ($\lambda=6$), coherence between different wells persists since $\vert g^{(1)}(x,x') \vert^2$ is significantly larger than zero at off-diagonal values $x\neq x'$ for all tilts, Fig.~\ref{Fig:3}(a)--(d).
For larger interaction strengths ($\lambda=20$), inter-well coherence is absent for no tilt ($\alpha=0$), Fig.~\ref{Fig:3}(e). 
As the tilt increases, inter-well coherence between populated neighboring wells is gradually restored, see Fig.~\ref{Fig:3}(a)--(d) for $\lambda=6$ and Fig.~\ref{Fig:3}(f)--(h) for $\lambda=20$: $\vert g^{(1)}(x,x')\vert$ gradually grows towards unity for values $x\neq x'$. We note that a tilt-driven localization takes place for larger tilts, cf. white areas in Fig.~\ref{Fig:3}(c),(d),(h) and Fig.~\ref{Fig:1}(b).

The effect of interactions is to merely diminish the inter-well coherence, cf. Fig.~\ref{Fig:3} (a)--(d) and (e)--(h): the value of $\vert g^{(1)}(x,x')\vert^2$ is generally closer to unity on the off-diagonal $x\neq x'$ for small interactions [Fig.~\ref{Fig:3} (a)--(d)] as compared to larger interactions [Fig.~\ref{Fig:3} (e)--(h)].

We thus demonstrate that an increase of the tilt, at a fixed interaction strength, assists inter-well coherence of bosons in neighboring wells, while an increase of the interaction strength, for fixed moderate barrier heights, diminishes inter-well coherence. 

We now analyze the correlations for larger barrier heights ($V_0=180$), Fig.~\ref{Fig:3}(i)--(p). 
For zero tilt and in comparison to moderate barrier heights, inter-well coherence is completely lost at large barrier heights, $\vert g^{(1)} (x,x') \approx 0 \vert^2$ for $x\neq x'$ in Fig.~\ref{Fig:3}(i),(m).

By comparing the correlations at moderate barrier heights to the correlations at larger barrier heights, we find -- as expected -- that a larger value of $V_0$ increases the degree of localization of the system. This is true, independently of the interparticle interaction strength; compare first and third as well as second and fourth row of Fig.~\ref{Fig:3}.

As for moderate barrier heights, a restoration of coherence is also seen as the tilt $\alpha$ is increased for larger barriers $V_0$. However this restoration of coherence is of different nature than in the case of moderate barriers: for large barriers a revival of the next-to-nearest neighbors coherence is seen, while the nearest neighbors remain incoherent, see Fig.~\ref{Fig:3}(j) for weak interactions and Fig.~\ref{Fig:3}(o) for strong interactions. This is a qualitative difference to the case of moderate barrier heights, where the restoration of coherence as the tilt increases is limited to neighboring sites in the lattice.

The effect of stronger interactions is, one, to defer the restoration of coherence to larger tilts (from $\alpha=2.5$ for $\lambda=6$ to $\alpha = 6.5$ for $\lambda=20$) and, two, to shift the tilt-driven localization of the bosons to larger tilts. At $\alpha=16$ ($6.5$) two (three) wells are still populated for $\lambda=20$ as opposed to $\lambda=6$ where only two (one) well is populated.

We assess the generality of our findings for the coherence properties for long-range interactions in the Supplemental Material~\cite{SupplMat}. The inclusion of long-range interactions favors the fragmentation of the BEC for a larger barrier height. We find that our main conclusions for short-ranged interactions hold also for the case of long-ranged interactions.

\paragraph*{Conclusions} Our analysis has shown intriguing features of the first-order correlation and coherence of bosons in a tilted triple well. Given the ease in defining the system parameters in experimental setups with ultracold bosons, our work provides a protocol to manage the coherence of the many-body state: a variety of correlation patterns is accessible simply by appropriately choosing the interaction strength, potential depth and tilt. Superfluid states -- associated with condensation -- can be created either localized in one well or delocalized across all wells. Mott-insulating states -- associated with fragmented systems -- with a customized particle number imbalance between distinct wells can also be prepared. The superfluid fully coherent state and the Mott-insulating fully incoherent phase represent extreme cases. Fig.~\ref{Fig:3} illustrates how intermediate degrees of correlation can also be achieved. The counter-intuitive revivals of coherence between next-to-nearest neighboring sites seen in panels (j) and (o) of Fig.~\ref{Fig:3} hint that even a management of non-local correlations is possible, if the control on the tilt and interaction strength is sufficiently accurate.

\paragraph*{Acknowledgements} Computational time at the HLRS Hazel Hen cluster is gratefully acknowledged. MCT acknowledges FAPESP for financial support and Hans Kessler for useful discussions. AUJL acknowledges financial support by the Austrian Science Foundation (FWF) under grant No. F65 (SFB ``Complexity in PDEs''), and the Wiener Wissenschafts- und TechnologieFonds (WWTF) project No MA16-066 (``SEQUEX'').


\begin{thebibliography}{9}
\bibitem{units} To define dimensionless units, we divide the Hamiltonian by $\hbar^2/(2mL^2)$, where $m$ is the mass of the considered boson and $L$ a convenient length scale. We first choose a length scale of $L=1\mu$m. The scale of energy for the mass of $^{87}Rb$ is $\hbar^2/(mL^2)=2\pi \hbar \times 116$Hz and the scale of time is $mL^2/\hbar=1.37$ms. The one-dimensional scattering parameter $\lambda$ is related to the three-dimensional scattering length $a_{3D}$ by $\lambda = 2Lm\omega_\perp a_{3D}/\hbar$, where $\omega_\perp$ is the frequency of the transversal confinement~\cite{olshanii}. Using $a_3D = 100.4a_0$ , where $a_0$ is the
Bohr radius, and $\lambda=6$ ($\lambda=20$), one obtains $\omega_\perp=41.3$kHz ($\omega_\perp=1376$kHz).
\bibitem{num_details} We set $f_{w}(x)=\Big(\frac{x}{x_{c}}\Big)^{60}$ and $x_{c} = \frac{3\pi}{2k}-\frac{1}{k}$ to
define the effective extension of the triple well. Our numerical grid runs from $x_{min}=-x_{c}-{1.09}/{k}$ to
$x_{max}=x_{c}+{1.09}/{k}$ and we use $256$ grid points.
\end{thebibliography}
\end{document}